\documentclass[sn-apa]{sn-jnl}

\usepackage{enumitem}
\usepackage{float}
\usepackage{hyperref}
\usepackage{color,soul}

\jyear{2024}%

\theoremstyle{thmstyleone}%
\theoremstyle{thmstyletwo}%

\theoremstyle{thmstylethree}%

\algnewcommand\Input{\item[\textbf{Input:}]}%
\algnewcommand\Output{\item[\textbf{Output:}]}%

\begin{document}

\title[Facilitating Interdisciplinary Knowledge Transfer]{Facilitating Interdisciplinary Knowledge Transfer with Research Paper Recommender Systems}


\author*[1,2]{\fnm{Eoghan} \sur{Cunningham}}\email{eoghan.cunningham@insight-centre.org}

\author[1,2]{\fnm{Barry} \sur{Smyth}}\email{}

\author[1,2]{\fnm{Derek} \sur{Greene}}\email{}

\affil*[1]{\orgdiv{School of Computer Science}, \orgname{University College Dublin}, \orgaddress{\city{Dublin}, \country{Ireland}}}

\affil[2]{\orgdiv{Insight Centre for Data Analytics}, \orgaddress{\city{Dublin}, \country{Ireland}}}



\abstract{In the recommender systems literature, novelty and diversity have been identified as key properties of useful recommendations. However, these properties have received limited attention in the specific sub-field of research paper recommender systems. In this work, we argue for the importance of offering novel and diverse research paper recommendations to scientists in order to reduce siloed reading, break down filter bubbles, and promote interdisciplinary research. We propose a novel framework for evaluating the novelty and diversity of research paper recommendations that leverages methods from network analysis and natural language processing. Using this framework, we show that the choice of representational method within a larger research paper recommendation system can have a measurable impact on the nature of downstream recommendations, specifically on their novelty and diversity. We describe a paper embedding method that provides more novel and diverse recommendations without sacrificing precision, compared to other state-of-the-art baselines.}




\maketitle

\section{Introduction}\label{sec:introduction}

As the rate of scientific publication continues to accelerate, it is increasingly difficult for researchers to keep abreast of developments in their own fields of study, and even more difficult to follow advancements in areas beyond their primary discipline. Many now rely on research paper recommender systems (RP-Rec-Sys), which combine ideas from recommender systems \citep{ricci2015recommender}, user modeling personalisation \citep{ghorab2013personalised}, information retrieval and discovery \citep{liu2019data}, to support their reading. Various approaches have been adopted for different discovery use-cases \citep{bai2018scientific}, with each use-case advocating a unique approach to recommendation. For example, \cite{hua2020path} and \cite{hui2020hybrid} consider how researchers engage with a new topic for the first time, whereas \cite{sharma2020predictive} focus on subject matter experts conducting literature reviews.

A common challenges across all applications of RP-Rec-Sys is how to prevent the so-called \emph{filter-bubble} effect of conventional recommender systems \citep{portenoy2022bursting}. This can reinforce siloed reading and citation tendencies by prioritising items that are similar to past recommendations. Amid efforts to understand and encourage boundary-crossing research \citep{qi2024facilitating}, recommendations that are over-reliant on similarity can be antithetical to the needs of modern interdisciplinary science, where the most valuable breakthroughs often come from surfacing the latent or unexpected connections between disciplines \citep{shi2023surprising,uzzi2013atypical,bornmann2019we}. Over the last 20 years, Rec-Sys evaluations have evolved to combat the filter bubble effect with the development of metrics like `recommendation novelty' and `recommendation diversity' \citep{smyth2001similarity,castells2021novelty}. However, to date, these evaluation perspectives have not been considered in the context of RP-Rec-Sys  \citep{ali2021overview}. In particular, three recent surveys of the RP-Rec-Sys literature highlight recommendation `diversity', `novelty', and/or `serendipity' as open challenges that have not been addressed sufficiently \citep{kreutz2022scientific,bai2018scientific,ali2021overview}. Addressing this gap in the literature is important as \textit{relevant}, \textit{diverse} and \textit{novel} recommendations (as they are defined by the Rec-Sys community), would serve to expose researchers to relevant works from outside of their established field. Such trans-disciplinary recommendations would be of great value given the importance and impact of interdisciplinary research \citep{shi2023surprising,lariviere2015long,chen2022interdisciplinarity}.


Motivated by the above, this work proposes a novel framework for evaluating research paper recommendations. The framework is designed to bridge the divide between RP-Rec-Sys evaluations and the broader Rec-Sys literature, and to facilitate and promote far-reaching, cross-discipline research recommendations and interdisciplinary knowledge transfer.
Specifically -- in addition to established measures of recommendation relevance -- we employ novel metrics for research paper recommendation \textit{diversity} and \textit{novelty} (or \textit{unexpectedness}) according to citation network distances and research topic dissimiliarities. 
We demonstrate our proposed framework with an experiment where we evaluate four recommenders employing different representation (or \textit{paper embedding}) methods. We show that the choice of paper embedding method upstream of an RP-Rec-Sys has a measurable impact on both the quality (as measured by relevance) and interdisciplinary (as measured by diversity and novelty) of downstream recommendations. 
In particular we highlight a recent embedding method --  ComBSAGE \citep{cunningham2023graph} -- that provides more far-reaching, interdisciplinary recommendations without compromising relevance when compared to other state-of-art approaches.

The remainder of this work is structured as follows. Section \ref{sec:background} formalises the tasks of research paper recommendation and embedding, and outlines important existing works in the area. 
Section \ref{sec:method} explains the framework we propose for evaluating research paper recommender systems. In particular, we emphasise our approach for measuring diversity and novelty in the context of RP-Rec-Sys. In Section \ref{sec:experiment_design}, we demonstrate our proposed framework to investigate the effect of research paper representation on downstream recommendations in RP-Rec-Sys. First, we summarise the dataset compiled for our analysis in Section \ref{sec:dataset}. Second, 
we detail the implementation of the paper embedding methods that we evaluate. Third, we describe our approach for generating recommendations given these article representations, and finally, the evaluation metrics used. The results of our evaluations are presented in Section \ref{sec:results} and we offer some final conclusions in Section \ref{sec:conclusions}.

\section{Background}
\label{sec:background}

We begin by formalising the task of research paper recommendation in Section \ref{sec:background_recommendation} where we outline several important aspects of RP-Rec-Sys evaluations, before discussing the concepts of \textit{diversity} and \textit{novelty} as they relate to both science-of-science and recommender systems in Section \ref{sec:diversity}. Finally, we discuss research paper representation learning (or embedding) in Section \ref{sec:background_representation}, as it is a crucial aspect of RP-Rec-Sys and the subject of our experiments.

\subsection{Research Paper Recommendation}
\label{sec:background_recommendation}

Research paper recommender systems (RP-Rec-Sys) are designed to suggest research papers that align with the interests or the needs of an end-user.
Consistent with the field of Rec-Sys and its broader applications, approaches to research paper recommendation can be categorised into two groups: Content-Based Filtering (CB) and Collaborative Filtering (CF). The former methods involve generating recommendations for users based on descriptions or representations of items \citep{lops2011content,smyth2007case}, whereas CF methods primarily depend on interactions between users and items to provide recommendations \citep{koren2021advances,shi2014collaborative}. Many systems have been implemented from a variety of task-oriented perspectives -- recommending papers to authors \citep{jiang2023taprec}, recommending authors to authors \citep{portenoy2022bursting}, and even recommending citations for papers \citep{ali2021overview}. Within each of these perspectives, many possible use-cases of RP-Rec-Sys exist. For example, \cite{hua2020path} and \cite{hui2020hybrid} focus specifically on researchers engaging with a new subject for the first time, while \cite{sharma2020predictive} develop their system to support researchers conducting in-depth literature review. All CB methods of RP-Rec-Sys (and many CF/hybrid methods), rely on some fundamental underlying form of paper representation or embedding. 

The diverse approaches to RP-Rec-Sys are reflected in the broad range of evaluation methods proposed in the literature.
One of the most reliable means of evaluating RP-Rec-Sys is through user studies: users interact directly with the system, and their satisfaction is assessed either through explicit surveys or deduced from factors like session duration  \citep{portenoy2022bursting}. Although these so-called `online' evaluation methods can be highly informative, they come with considerable costs, particularly in terms of recruiting and surveying users. Therefore, `offline' evaluations are often preferred as a more feasible option or employed as a preliminary step before proceeding with online evaluations. Offline evaluations typically require some (usually external) ground truth used to define which papers are relevant to a given user or query paper. In the context of scientometrics, existing relationships which have been used as ground truths include `paper-cites-paper' \citep{ali2022sprsmn}, `author-cites-paper' \citep{jiang2023taprec}, and `paper-co-read-with-paper' \citep{singh2023scirepeval} information. Of course, these choices will depend on the specific use-case and availability of appropriate data.

The choice of ground truth in RP-Rec-Sys evaluations has been identified as a core challenge in the domain \citep{ali2021overview}. For example, in those cases of RP-Rec-Sys where the user/author view is not considered, direct citation information (`paper-cites-paper') is commonly used as a ground truth for relevance. However, implementing such an evaluation while respecting the inherent temporal organisation of a citation network can be challenging.  Most methods involve excluding certain references from the bibliography of some query or test papers, and require the system to predict these citations. This process resembles the standard link prediction problem frequently posed in the field of network analysis. Consequently, only the most recent publications can be considered as queries. Otherwise, a user is tasking the model with predicting early citations given the knowledge of later citations, thus disregarding the real temporal ordering of the data. Avoiding this temporal data leakage limits evaluations to testing a single use-case of RP-Rec-Sys: `complete the bibliography of this unpublished research paper'. 

Of course, there are several other use-cases of RP-Rec-Sys that should be also considered, and it is important to be able to evaluate systems in such scenarios. Co-share and co-view relationships between papers have been employed as alternatives to direct citation \citep{singh2023scirepeval}. Such information is gathered by services like Semantic Scholar and can offer a highly valuable indication of relatedness or similarity between papers. However, the scope of any evaluations that rely on these user-based relations will be restricted to the small sample of papers for which they are available. Moreover, we have concerns that encouraging models to fit to this signal may only serve to reinforce established reading and citation behaviours. Given the above limitations of these popular methods, we rely on \textit{co-citation} relations in our evaluations as a ground truth for establishing relevance between two papers. Further details about our evaluation framework and its advantages can be found in Section \ref{sec:method}.

Regardless of the choice of ground truth, (e.g. `paper-cites-paper', `paper-co-saved-with-paper'), most evaluations employ metrics designed to measure the proportions of recommendations that are relevant to the user or to their query (e.g. precision), the quality of the ordering of the recommendations (e.g. DCG, MRR), the proportion of possible relevant papers that were recommended (e.g. recall), or some combinations of these metrics (e.g. MAP, F1). Exploring these different metrics may permit the system to be evaluated with respect to a small number of different use-cases. For example, a researcher who makes regular use of an RP-Rec-Sys may value high precision for small recommendation sets (i.e., the top $5$--$10$ papers), while an author undertaking some large survey or literature review may prefer high recall on a larger recommendation set (30+ papers). However, limiting the scope of evaluations to these metrics can exclude other valid RP-Rec-Sys use-cases from consideration. To date there has been limited attention directed towards recommendation \textit{diversity} and \textit{novelty} \citep{bai2018scientific,ali2021overview}, despite the emphasis that has been placed on these objectives in the broader recommender systems literature over many years \citep{smyth2001similarity,castells2021novelty}. Systems that perform well with respect to these metrics may facilitate another use-case -- authors seeking exposure to novel ideas, challenges, and methodologies from other research domains.


\subsection{Novelty and Diversity}
\label{sec:diversity}

Novelty and diversity are important concepts in both recommender systems research and scientometric analysis. As \citet{castells2021novelty} note, \emph{``novelty can be generally understood as the difference between present and past experience whereas diversity relates to the internal difference within parts of an experience''}. Each concept has been formalised and understood independently in these two fields, and while usage of these concepts differs somewhat across the two domains, both perspectives are closely related and remain relevant to this work. In the following section, we will outline the concepts of novelty and diversity as they relate to scientometric analysis and recommender systems, and crucially, how \textit{research} novelty and diversity may be supported and encouraged via \textit{recommendation} novelty and diversity.  

In the field of scientometrics, a research output may be deemed \textit{novel} if it involves a `surprising' or `unlikely' integration of ideas, methods, or models \citep{shi2023surprising, shin2022scientific}. It has been shown that the most impactful and ground-breaking advancements are often led by works which maximise this definition of novelty, and thus research novelty is highly valued. Diversity has been considered and evaluated from a number of similar perspectives. For example, many studies measure the diversity of topics, disciplines, or subject categories referenced within an article, using this as a metric to gauge the intellectual breadth of a single work. \citep{porter2007measuring,okamura2019interdisciplinarity,lariviere2015long}. Others evaluate this across a researcher's entire body of work to determine the breadth of their research diversity \citep{porter2007measuring,yu2023analyzing}. In this manner, research diversity serves as an indicator for multi- and interdisciplinary work, which has been shown to correlate with research impact \citep{lariviere2015long,chen2022interdisciplinarity}. 

In the area of recommender systems, diversity and novelty take on different interpretations. For the purposes of this work, we will consider the following formulations of the \textit{diversity} and \textit{novelty} of a set of recommendations \citep{castells2021novelty}. For a set of $k$ recommendations given for some query, we denote the novelty of the set of recommendations as the mean cosine distance between the representation of a recommended item and that of the query. 
Similarly, the diversity of the set of $k$ recommendations can be defined as the mean pairwise cosine distance between the representations of recommended items. This is sometimes referred to as the Average Intra-List Distance \citep{smyth2001similarity,castells2021novelty}. 
Both concepts are considered desirable across many recommender systems applications \citep{vargas2011rank,kaminskas2016diversity,nakatsuji2010classical,zhang2013definition}. For example, enhancing recommendation diversity can be a means of reducing the redundancy of 
a set of recommendations, as is the case with travel recommender systems, which usually seek to avoid suggesting similar properties in the same destination resort \citep{smyth2001similarity}; in this case, recommending a broader set of options is an important way for a recommender system to `hedge its bets' especially when the precise needs and preferences of the user are uncertain. Likewise, when recommending movies or music to consumers, novelty is often desirable according to the inherent satisfaction of unexpectedness, change, and complexity, and also as a foil to satiation and the decreased satisfaction associated with repeated consumption \citep{castells2021novelty,nakatsuji2010classical}.

When recommending research papers, it is crucial to recognise the difference between the respective formulations of \textit{research} novelty and diversity, and \textit{recommendation} novelty and diversity. We consider a recommended research paper to be a \emph{novel recommendation}, if it is unexpected with respect to the user or query (as per Equation \ref{eq:novelty}), and not simply if the recommended paper represents novel work according to definitions of \citep{shi2023surprising} or \citep{shin2022scientific}. Of course, the former does not exclude the latter -- a recommendation may be novel according to both definitions. In fact, it follows that suggesting recommendations of high \textit{recommendation} novelty could increase the \textit{research} novelty of a researcher's future outputs, as they go on to integrate the ideas from surprising or otherwise distant sources. We argue that the same is true for research diversity. A set of recommended papers with high \textit{recommendation diversity} may not contain any works which are themselves diverse or interdisciplinary according to existing scientometric definitions \citep{porter2007measuring,okamura2019interdisciplinarity}. However, a crucial motivation for increasing \textit{recommendation} diversity in RP-Rec-Sys is to promote \textit{research} diversity in future works. 

To date, while recommendation diversity, novelty, and the related concepts of `serendipity' and `surprise' have been explored by the wider recommender systems community \citep{castells2021novelty}, they are less well understood in the context of research paper recommender systems. For instance, in a recent survey by \citet{ali2021overview} of 67 models for RP-Rec-Sys (published between 2013--2021), only 3 papers were found to evaluate recommendation `novelty'. Other surveys have highlighted diversity, novelty, and serendipity as key challenges \citep{bai2018scientific,kreutz2022scientific} in the domain and high-priority future work. 

Diverse and novel recommendations are particularly important in the field of research paper recommendation, to broaden the horizons of researchers, by exposing them to novel ideas, methodologies and challenges, usually from other fields, and by ultimately removing barriers to interdisciplinary research. This is especially valuable given that interdisciplinary research outputs, which integrate ideas from several distinct disciplines have been consistently found to be more impactful that the more conventional, incremental, within-discipline outputs \citep{shi2023surprising,okamura2019interdisciplinarity,lariviere2015long}. 

Existing works that evaluate research paper recommendation diversity and novelty, are limited and varied. \cite{nishioka2019research} attempted to use additional information from authors' \textit{twitter} profiles to provide novel or `serendipitous' recommendations, and relied on a small user-study as an `online' evaluation. Other studies have developed offline evaluations, for example: \cite{rodriguez2019discovering} measured recommendation serendipity, according to how a set of recommendations differ from a set of recommendations provided by some `primitive' model, while \cite{chaudhuri2021hidden} assessed the \textit{research} or \textit{disciplinary} diversity of individual papers recommended, rather than the \textit{recommendation} diversity of the set of recommended articles. 

\subsection{Scientific Document Representation}
\label{sec:background_representation}

\begin{figure}[h]%
\centering
\vskip 0.5em
\includegraphics[width=\textwidth]{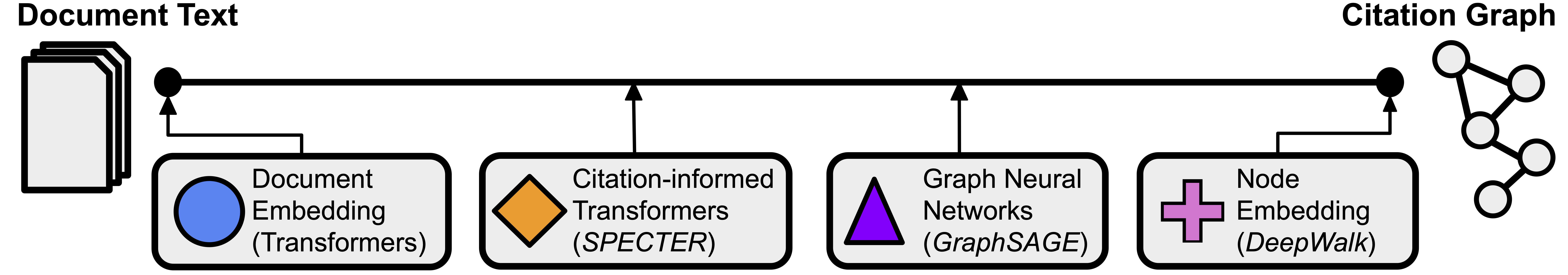}
\caption{The spectrum of scientific document representation methods. Methods for paper embedding can be positioned along this spectrum, with those methods exclusively reliant on article text placed on the left, and those methods exclusively reliant on the citation graph placed on the right.  Architectures that make use of both sources of information are situated between these two extremes.}
\label{fig:spectrum}
\end{figure}

Representation learning for scientific documents refers to the task of representing research papers in a vector space so that the important relationships between the papers are preserved -- i.e., 
semantically related papers should have similar representations (or \textit{embeddings}) \citep{kozlowski2021semantic}. These research paper embeddings can then be used to support downstream tasks such as literature review, classification, or (as is the focus of this work) recommendation. Scientific articles can be related in different ways. For example, two papers may be related if they have similar content, or two papers might be related if there is a transfer of knowledge or ideas between them; that is, there is some citation or chain of citations from one to the other \citep{jo2022see}. Accordingly, article text and citation relationships are the main sources of information for scientific document representation \citep{kozlowski2021semantic}. Some approaches to document embedding rely on additional article metadata, such as information about the authors or the publication venue. We exclude this information from our discussion of paper embedding as we believe such metadata can surface relations between papers that are only superficial in the context of some downstream use-cases. For example, in the RP-Rec-Sys use-case of exposing researchers to relevant works from outside the scope of their current reading patterns, author or venue relations may only serve to reinforce their existing behaviours. Of course, many use-cases exist where authorship information is relevant (e.g. seeking works written by some author) or even vital (e.g. personalised recommendations based on an author's bibliography). In such cases, methods have been developed for including these relations downstream in the recommender system and not in the article embedding stage \citep{ali2022sprsmn,mei2022mutually,jiang2023taprec}. Thus, while additional relations based on metadata can be employed downstream according to the use-case or application, we prefer to rely only on article content and citation relations during the article embedding stage. We focus our analysis on methods that make use of these two modalities, and propose that all of the approaches we include can be considered on the spectrum of techniques illustrated in Figure \ref{fig:spectrum}. We position methods that rely only on the text of an article on the left-hand side, those that rely only on citations on the right, while hybrid methods (combing text and citations) lie in between these extremes. 

Simple text-based approaches to research paper representation learning (including those applied downstream to recommendation tasks) use 'bag-of-words' vectors, which count the occurrence of each word in each document such that documents which contain similar vocabularies will have similar representations. These approaches can be extended to re-weight or transform the features according to a features utility in discriminating between documents relating to different topics (e.g. TF-IDF, LDA, LSA, etc.) \citep{achakulvisut2016science, amami2016lda, dai2018joint}. Transformer-based language models, such as BERT \citep{devlin2018bert} (Bidirectional Encoder Representations from Transformers), leverage attention mechanisms and self-attention layers to transform the textual content of research papers into document embeddings \citep{nogueira2020navigation}. Crucially, models like \textit{SciBERT} \citep{beltagy2019scibert} have been pre-trained on large corpora of research papers in order to capture the specific vocabulary and semantics of academic writing, and are commonly applied to an article's title and abstract to produce paper embeddings. 

At the other end of the spectrum, some methods exclusively depend on information from the citation network. Given a graph $G = (V,E)$, where $V$ is the set of nodes or vertices on the graph, and $E$ is the set of edges, or interactions between those nodes, 
\textit{Graph Representation Learning} refers to the task of learning latent features/low-dimensional vectors to describe nodes on the graph $G$. These \textit{node embeddings} offer a compact and informative representation of the nodes which preserve important properties, such as proximities and community structure in the graph. In the case of a network of research papers, where nodes represent research articles and the interactions between the nodes represent citation relations for example, node embeddings can used as the source of paper embedding. Thus, papers which are nearby to each other in the citation network, (e.g. they share many citations or references), will have similar representations. In addition to the purely network-based methods for article embedding (e.g. \textit{DeepWalk}) and the purely NLP-based approaches (e.g. \textit{Sci-BERT}), the complex nature of research papers and their relations necessitates approaches that are capable of leveraging (and combining) both modalities, to provide a single article representation. 

Na\"ive approaches have been developed which simply combine or concatenate text and network embeddings. However, more sophisticated techniques have been suggested to effectively merge the semantic and relational dimensions of research articles. For example, \textit{SPECTER} \citep{cohan2020specter} uses citation relations between papers as an external inter-document signal to fine tune text-based representations. Initialised with \textit{SciBERT} and using title and abstract text from articles, \textit{SPECTER} is trained to minimise the distance between the representations of a query paper and an article that it cites, while maximising the distance between the query paper and an article that it does not cite. Thus, in addition to the vocabulary and semantics of a research article's content, \textit{SPECTER} uses the knowledge transfer described by citation relations to inform similarities in the embedding space, making it a good choice for downstream RP-Rec-Sys applications \citep{ali2022sprsmn}.

Message Passing Graph Neural Networks (MP-GNNs) (e.g. GraphSAGE, GCN, VGAE) can be applied to graph structured data, where some additional information is available for the nodes. 
Given a graph, $G = (V,E)$ with node features $X$, 
one layer of an MP-GNN is typically described with two functions: \textit{aggregate} and \textit{update}. For a node $v_i$, with features described by a vector $x_i$, we denote the neighbours of $v_i$ as $\{v_j,v_k, ... ,v_p\}$ with features $\{x_j,x_k, ... ,x_p\}$. An embedding for $v_i$ (denoted by $h_i$) is then generated via the update $h_i \gets \phi(x_i, u_i)$, where $u_i \gets \psi(\{x_j,x_k, ... ,x_p\}, x_i)$. Thus, $\psi$ is some (typically permutation invariant) function for \textit{aggregating} information from neighbouring nodes (potentially with consideration of $x_i$), and $\phi$ is some parametric function for \textit{updating} the representation of a node using the aggregated message. MP-GNNs can be trained in a supervised fashion (e.g. in pursuit of some task such as node classification), or in an unsupervised manner where the objective task used for training is graph reconstruction (or link prediction). 
MP-GNNs are designed to learn node embeddings which depend not only on a node’s descriptive features, but also the features of its neighbours. MP-GNNs iteratively update the representation of each node by receiving information from neighbouring nodes and learning aggregation functions to combine those messages with a node’s current representation. In the context of a citation network, where a node’s features are typically derived from the text of the corresponding paper, the representation of some focal paper is informed not only by its own content, but also by the content of the papers that it cites (the ideas upon which it develops), and even by the content of the articles that cite the focal paper (its applications or the work that builds on it). For these reasons MP-GNNs are commonly applied to document representation and RP-Rec-Sys \citep{ali2021overview,kreutz2022scientific,kozlowski2021semantic,jiang2023taprec}.


\section{Method: Evaluating Research Paper Recommendations}
\label{sec:method}
In this section, we present the primary technical contribution of our work: a novel approach to RP-Rec-Sys evaluations. Our framework draws on contemporary practices in the wider Rec-Sys literature, and aims to consider recommendation novelty and recommendation diversity from multiple perspectives. Additionally, our method is intended to avoid a number of common pitfalls associated with RP-Rec-Sys evaluations. Specifically, it is designed to be robust and consistent with the many diverse use-cases of RP-Rec-Sys and crucially, to respect the inherent temporal ordering of a citation network.
Our proposed method can be considered in two stages. First, we describe our approach for offline evaluation of RP-Rec-Sys \textit{relevance}. Secondly, we present four metrics for evaluating RP-Rec-Sys \textit{novelty} and \textit{diversity}.

\subsection{Evaluating Recommendation Relevance} 
Given a query paper $q$, we first evaluate a set of recommendations $r$, according to their relevance. We adopt co-citation relations as our ground-truth for relevant papers. Formally, a recommended paper is relevant to the query paper if and only if both papers are co-cited in some third publication. Co-citation relations have been widely used as a measure of relevance in scientometric analyses \citep{small1973co,chen02proc,gmur2003co,zhang2022citation}. A co-citation may occur between two papers for a number of reasons; (i) an author lists/compares/evaluates two papers that are methodologically/conceptually similar; (ii) an author outlines an important transfer of knowledge from one paper to the other; (iii) an author highlights an interesting problem and a potential solution. 

Crucially, co-citation between papers is a strong and flexible signal for relation that is established and supported by the authors active in the area. Moreover, in many applications, co-citation may prove a more appropriate ground truth compared to other paper-paper relations (such as those outlined in Section \ref{sec:background_recommendation}). For example, direct citation between papers is often implemented as the ground truth for relevant recommendations \citep{mei2022mutually,nogueira2020navigation,ali2022sprsmn}. However, it is challenging to evaluate such a system while still adhering to a strict temporal split and avoiding any potential data leakage. As direct citations can only occur at the time of publication, the set of query papers in evaluations that rely on direct citations must be restricted to contain only unpublished articles. This specific case of RP-Rec-Sys is commonly referred to as `citation recommendation' \citep{ali2021overview}. Direct citation ground truths should not be applied outside of this use-case, as they disobey the inherent temporal structure of a citation network. As such, we propose the use of co-citation relations to allow users to query systems with published research, such as examples of their own works, (i.e., for personalised recommendations), or queries relevant to their ongoing works (i.e., for information retrieval or supporting literature reviews).

Accordingly, the relevance of a set of recommendations can be measured using simple metrics such as \textit{precision}, the proportion the articles recommended in $r$ that are relevant to $q$, and \textit{recall}, the proportion or articles relevant to $q$ that are recommended in $r$. Alternatively, relevance could be measured using more complex metrics which consider the ordering of recommended articles within the set $r$, such as Area Under the ROC Curve (AUC) or Discounted Cumulative Gain (DCG).

\subsection{Evaluating Recommendation Diversity and Novelty}
\label{sec:measuring_diversity_and_novelty}
In addition to understanding the relevance of recommended articles, it is pertinent in many use-cases to also evaluate the diversity and novelty of a set of recommendations. Diverse and novel recommendations have many benefits across different applications (see Section \ref{sec:background_recommendation}). In the specific domain of research paper recommender systems, diverse and novel recommendations (that remain relevant to the user) can expose researchers to otherwise unknown methods/solutions/challenges from distant fields of study. In this way, RP-Rec-Sys can promote the exchange of ideas across disparate disciplines and burst scientific filter bubbles. \citep{portenoy2022bursting}

To assess the diversity and novelty of the recommendations generated, we present four metrics: (i) citation network recommendation novelty, (ii) citation network recommendation diversity, (iii) article content recommendation novelty, and (iv) article content recommendation diversity. In our calculations, we use the widely accepted, standard formulations of novelty and diversity \citep{castells2021novelty,smyth2001similarity} as presented in Equations \ref{eq:novelty} and \ref{eq:diversity} below. 

For a set of $k$ recommendations $r$ provided for some query $q$, we denote the novelty of the set of recommendations as
\begin{equation}
\label{eq:novelty}
\text{recommendation novelty} = \dfrac{1}{k} \sum_{i \in r} \dfrac {x_q \cdot x_i} {\left\| x_q\right\| _{2}\left\| x_i\right\| _{2}}
\end{equation}
which is the mean cosine distance between the representation $x_i$ of a recommended item $i$ and that of the query. 
Similarly, the diversity of the set of $k$ recommendations can be defined as 
\begin{equation}
\label{eq:diversity}
\text{recommendation diversity} = \dfrac{1}{k} \sum_{i,j \in r} \dfrac {x_i \cdot x_j} {\left\| x_i\right\| _{2}\left\| x_j\right\| _{2}}
\end{equation}
which is the mean pairwise cosine distance between the representations of the recommended items.

We consider diversity and novelty from two perspectives: the citation network perspective, and the article content perspective, thus giving four metrics. 
Specifically, when a system is posed with a query and provides a set of recommendations, we measure:
(i) citation network recommendation \textit{novelty} as the mean cosine distance between the \textit{DeepWalk} representation of the query, and the \textit{DeepWalk} representation of the recommendations, such that recommendations are considered `novel' if they are positioned far away from the query in the citation network;
(ii) citation network recommendation \textit{diversity} as the mean cosine distance between the \textit{DeepWalk} representation of the query, and the \textit{DeepWalk} representation of the recommendations, such that a set of recommendations are considered `diverse' is they are positioned far away from each other in the citation network;
(iii) article content recommendation \textit{novelty} as the mean cosine distance between the \textit{SciBERT} representation of the query, and the \textit{SciBERT} representation of the recommendations, such that recommendations are considered `novel' if they have content that differs from the query;
(iv) article content recommendation \textit{diversity} as the mean cosine distance between the \textit{SciBERT} representation of the query, and the \textit{SciBERT} representation of the recommendations, such that recommendations are considered `diverse' if they have content that differs from each other.

\begin{figure}[!t]%
\centering
\includegraphics[width=\textwidth]{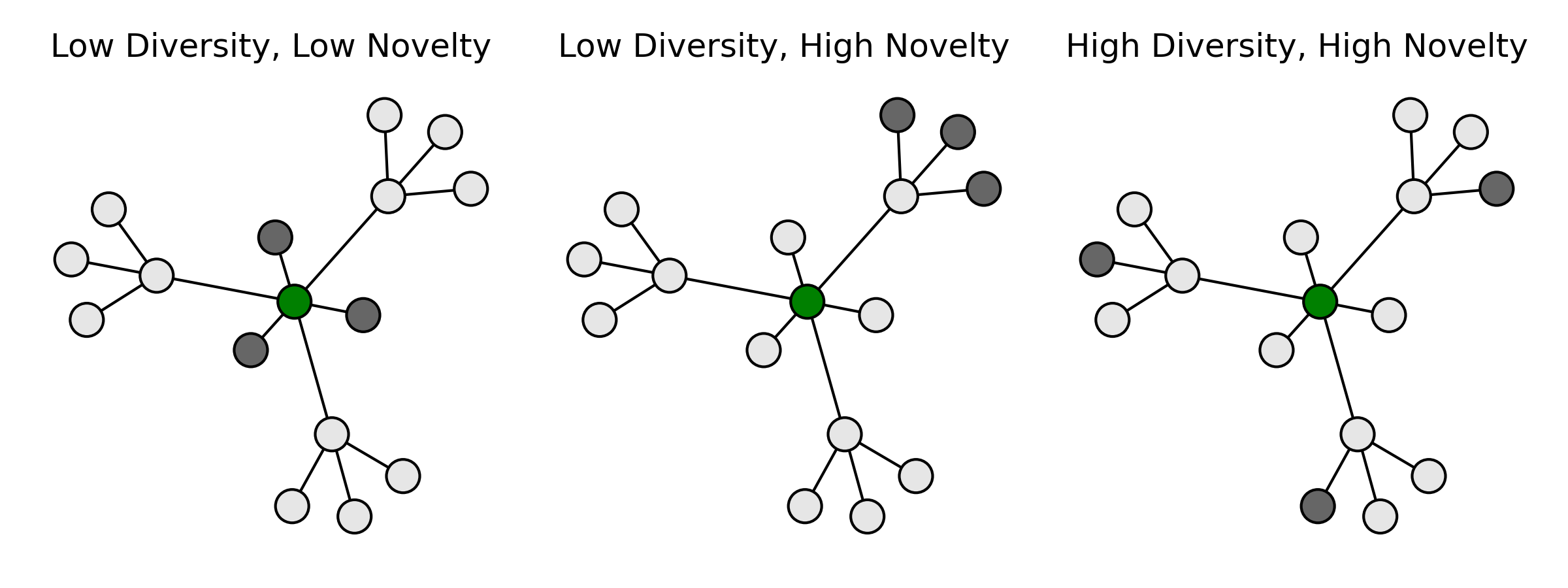}
\caption{A toy example designed to offer a visual demonstration of the network perspective of novelty and diversity in recommendation. The central node represents a query and the shaded nodes show examples of recommendations with different diversity and novelty scores, as measured using equations \ref{eq:novelty} and \ref{eq:diversity}, given \textit{DeepWalk} embeddings as descriptions of nodes.}
\label{fig:network_perspective}
\end{figure}

Metrics (i) and (ii) above (diverse and novel recommendations according the citation graph perspective) are illustrated in the toy example provided in Figure \ref{fig:network_perspective}.  
According to our proposed metrics, a recommended paper can be considered to be a `novel' or `unexpected' recommendation if it is dissimilar to the query with respect to its position within the citation network (as in (i)), or its content (as in (iii)). These calculations of `recommendation novelty' are independent to the scientific or methodological novelty of the works recommended, (see the discussion in Section \ref{sec:background_recommendation}). 
Recommendation `diversity' measures how different the recommended papers are from one another. Thus, unlike recommendation novelty, recommendation diversity cannot be evaluated for a single recommendation. 

Similarly, the title and abstract textual content perspective considers distances between the \textit{SciBERT} embeddings of the title and abstract of the research papers when measuring diversity and novelty of recommendations. As discussed in Section \ref{sec:background_representation}, the \textit{SciBERT} embeddings that we use to represent the content of research papers correspond to the final hidden state representations of the [CLS] token in the large pre-trained BERT model, \textit{SciBERT}. 


Of course, the utility of novel and diverse recommendations is largely contingent on their relevance. As such, we also suggest calculating and reporting the novelty and diversity of the subset of the recommended papers that are \textit{relevant} to the query. We note that in some cases, the diversity of the set of recommended and relevant papers may not be a particularly meaningful statistic,  as it is often calculated from a set of papers that is substantially smaller than $k$. However, we include it here for completeness.

\section{Experiment Design}
\label{sec:experiment_design}
In this section we describe the design of an experiment to demonstrate our proposed evaluation methods and to investigate the potential for RP-Rec-Sys to facilitate interdisciplinary research and knowledge transfer. Specifically, we evaluate different recommenders -- each employing a different representation method -- to explore the effect of upstream research paper embedding methods on downstream recommendations in an RP-Rec-Sys context. We implement four different approaches to research paper embedding: one baseline, two state-of-the-art, and a recent method yet to be implemented in an RP-Rec-Sys context. These implementations are detailed in Section \ref{sec:experiments_embedding}. We generate recommendations according to these embeddings using a simple, representation-agnostic recommender system which is described in Section \ref{sec:experiments_recommendation}. Subsequently, we evaluate each set of recommendations according to the framework described in Section \ref{sec:method}. In particular, we consider the relevance, diversity, and novelty of recommendations. Section \ref{sec:experiments_evaluation} outlines the evaluation metrics we employ, and how these metrics can relate embedding methods to their most appropriate recommendation use-case. In addition to established use-cases, such as information retrieval and literature surveys, we focus on the use-case of exposing researchers to relevant challenges, methods, and ideas from works beyond their traditional disciplinary borders, in an effort to promote interdisciplinary knowledge transfer.

\subsection{Data}
\label{sec:dataset}
To conduct our experiment, we require a densely-connected citation network that includes research from a variety scientific disciplines. We constructed a novel citation graph using Semantic Scholar citation information for 58,513 research papers. To ensure that this graph contained regions of interdisciplinary research, we collected the set of research articles according to the following process:
\vskip 0.5em
\begin{enumerate}[label=\roman*]
    \item Select 8 topics from the All Science Journal Categorisation (ASJC): \emph{`Computer Science'}, \emph{`Mathematics'}, \emph{`Chemistry'}, \emph{`Medicine'}, \emph{`Social Sciences'}, \emph{`Neuroscience'}, \emph{`Engineering'}, and \emph{`Biochemistry, Genetics and Molecular Biology'};
    \item From each ASJC, select a random sample of up to 1,500 articles published in journals assigned to that topic;
    \item Collect any additional articles that have a citation relationship (`cited by' or `citing') with at least 10 articles in the seed set;
    \item Filter the resulting network by removing any articles for which abstract text is unavailable or in a language other than English, and retain only the largest fully connected component in the network.
\end{enumerate}
\vskip 0.5em
This process produced a densely-connected, multidisciplinary citation graph with title and abstract text for each article, and included 836,857 citations among the 58,513 research papers. The earliest paper was published in 1893 and the latest papers were published in 2022, but more than 96\% of the papers were published in the period 2000--2022. We used a language model that was pre-trained on a large scientific corpus to represent the text or articles as dense document vectors. Specifically, for each article, we encoded its concatenated title and abstract using the pre-trained SciBERT model \citep{beltagy2019scibert} and used the final representation of the [CLS] token to represent each document. The dataset is available at \url{https://doi.org/10.7910/DVN/P62Q5V}.


\subsection{Representations for Recommendation}
\label{sec:experiments_embedding}

We implement and evaluate four methods for research paper embedding: 
\vskip 0.5em
\begin{enumerate}
\item \textit{TF-IDF}, an established baseline for most document representation tasks;
\item \textit{GraphSAGE} \citep{hamilton2017inductive}, a scalable GNN architecture, commonly used for research paper representation learning and recommendation \citep{gao2023survey,jiang2023taprec,kozlowski2021semantic}; 
\item \textit{SPECTER} \citep{cohan2020specter}, an NLP-based approach also applied in downstream recommendation tasks \citep{ali2022sprsmn,singh2023scirepeval,church2024academic};
\item \textit{ComBSAGE} \citep{cunningham2023graph}, a novel approach to paper embedding based on the \textit{GraphSAGE} architecture. 
\end{enumerate}
\vskip 0.5em
For completeness, we now briefly outline the recent Community-Based Sample and Aggregation (ComBSAGE) \citep{cunningham2023graph} method for Message Passing Graph Neural Networks. The goal of this architecture is to incorporate local structural information in the message aggregation stage of MP-GNNs. This generates scientific document embeddings that are more appropriate for representing interdisciplinary research, without compromising the quality of paper representations overall.

\begin{figure}[h!]%
\centering
\includegraphics[width=\textwidth]{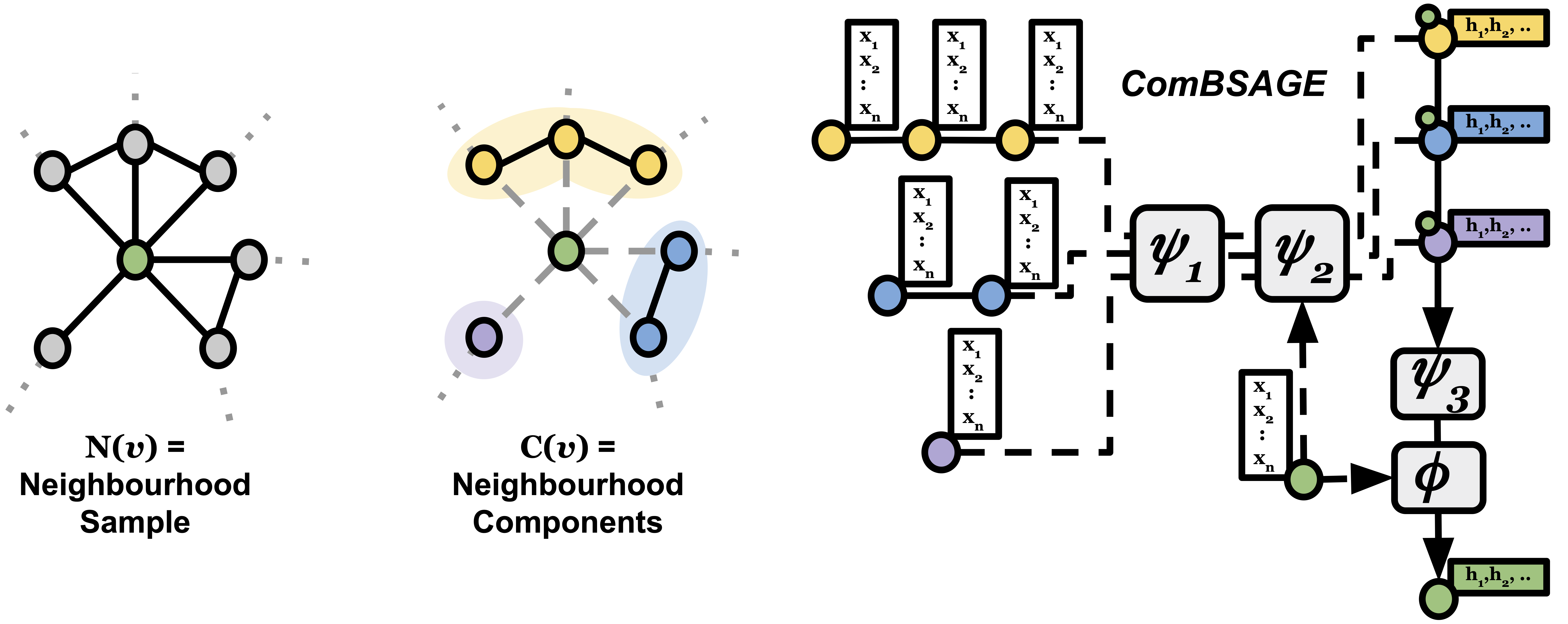}
\caption{One layer of the ComBSAGE MPGNN framework. Unlike traditional MPGNNs, messages are not aggregated uniformly across all neighbours. The neighbours of node $v$ are grouped by the neighbourhood component function $C$. Messages from each component in $C(v)$ are aggregated separately by $\psi_i$, $\psi_2$, before combination ($\psi_3$) and update ($\phi$). For full details on the architecture, see the workshop proceedings \citep{cunningham2023graph}.}\label{fig1}
\label{fig:method}
\end{figure}

The motivation behind this approach is as follows. Suppose we wish to aggregate messages from three neighbouring nodes, two of which are connected. We propose that these messages should be combined in a manner that accounts for this connection. Consider, for example, the task of representing a research paper which draws heavily on methods from the fields of mathematics and computer science, and that has a few recent applications in a field like political science. Such a paper presents in a citation network as a node with many connections to a large community of mathematics and computer science-related publications, and a smaller number of connections to works in political science. A traditional MP-GNN may lose much of the signal from the political science papers when aggregating messages. 
This is related to the \textit{over-smoothing} phenomenon, a known weakness in many existing GNN architectures, where node representations converge and become indistinguishable from their neighbours \citep{chen2020measuring}. It also connects to the \textit{over-squashing} problem, where information is lost in nodes that act as bottlenecks in the network \citep{topping2021understanding}. The proposed method of message aggregation splits messages from distinct communities, such that they can be aggregated separately before being combined (see Fig \ref{fig:method}). We suggest that this approach allows the important interdisciplinary implications of a research paper (e.g. its applications in political science) to be preserved in its representation, such that it may be recommended to or retrieved by authors in those secondary disciplines.

In the next section, we will outline the downstream RP-Rec-Sys framework that we apply to the chosen embedding methods, in an effort to characterise the differences in the recommendations that they provide. 

\subsection{Generating Recommendations}
\label{sec:experiments_recommendation}

To isolate the effect of the representation learning phase on downstream recommendations, we implement a simple recommendation framework (similar to \citep{bhagavatula2018content}), that can be applied independently of the upstream embedding method. Given a query paper $q$ from a set of research papers $P$, we seek to recommend the top $k$ relevant papers according to the paper descriptions or feature space $H$. That is, learn some function $g(h_q, h_i) = p( \text{recommend paper } i \mid \text{query paper } q)$, which computes the probability of recommending paper $i$, given query $q$ (or the probability that paper $i$ is relevant to query paper $q$). Thus, a set of recommendations is realised as the top $k$ papers which maximise $g$ for a given query.

To evaluate the recommendations produced by different systems, we divide the citation graph $G = (V,E)$ into discrete time steps. For a given year $y$, $G_{y} = (V_{y}, E_{y})$  contains all papers published up to $y$, $V_{y-2} \subseteq V_{y-1} \subseteq V_{y}$, similarly $E_{y-2} \subseteq E_{y-1} \subseteq E_{y}$. Thus, the citation graph for 2016 contains all papers published up to an including 2016, and any citations between those papers: $G_{2014} \subseteq G_{2015} \subseteq G_{2016}$. The paper metadata/content, as described by $X$, remains constant after publication. According to each paper embedding method, we train an embedding model $f$ on all data up to and including 2016. We then infer paper embeddings for all papers in that period: $f(G_{2016}, X) = H$, and train a multi-layer perceptron (MLP) recommendation function $g$, on any co-citing pairs where the co-citing paper is published no later than 2017. The MLP $g$ is trained to compute the probability of a co-citation occurring between a paper $i$ and query paper $q$, based on their respective embeddings $h_i$ and $h_q$. Accordingly, the probability of recommending a paper $i$, given query $q$ is given by
\begin{equation}
\label{eq:recommendation}
    p(\text{recommend paper } i \mid\text{query paper } q)  =  \sigma(W_2 \cdot a + b_2)
\end{equation}
where $a = \sigma(W_1 \cdot CONCAT(h_q,h_i) + b_1)$, $W_1, b_1$ and $W_2, b_2$ are the weights and biases of layers 1 and 2 of the MLP, and $\sigma$ is a non-linear activation function. The use of small (two-layer) feed-forward neural networks as the recommendation layer in larger RP-Rec-Sys is well established in the literature \citep{bhagavatula2018content,huang2015neural,li2019personalised}. Finally, we evaluate each approach to embedding using $f$ and $g$ to make recommendations for every paper published in 2017 (6211 papers). A recommendation is considered to be \textit{relevant} if the query paper and recommended paper are co-cited any time after 2017 (i.e., during $2017$--$2023$ in our data). Any co-citation between two papers according to Semantic Scholar is recorded, so we do not limit this relationship to only the co-citations which occur in the citation network we have collected. If there is a direct citation between the query paper and some co-cited paper, then we exclude that co-cited paper from the list of papers relevant to the query paper. This is done to avoid rewarding a system that makes recommendations directly from a query paper's bibliography or citations. We choose 2017 as the year for evaluation so that the citation network is sufficiently dense for training purposes, while also allowing the query papers to have sufficient time to accumulate citations, and thus co-citations/relevant papers.

\subsection{Evaluating Recommendations}
\label{sec:experiments_evaluation}

Using co-citation as a ground truth, we adopt two common metrics for assessing the relevance of a set of recommendations: \textit{precision} and \textit{recall}. For top-$k$ recommendations, precision$@k$ refers the proportion of recommended articles that are relevant, while recall$@k$ is proportion of the relevant articles that are recommended. In our later experiments, we report precision and recall at $k \in \{10, 20, 30\}$, where the lower end is chosen to represent the type of recommendations that might be regularly offered to a researcher/user (e.g. a short-list of recommendations in a weekly email), while the upper end could represent the case of an author conducting a literature review. Precision and recall scores are presented in Table \ref{tab:prec_rec}. Every paper published in 2017 is used a query paper, and we use bootstrapping to report means and standard errors. In practice, we consider recommendation relevance to be paramount across all RP-Rec-Sys use-cases. In particular, in a retrieval use-case, recommendation precision should be valued highly, while in a literature review setting, recommendation precision and recall would be the most important metrics. We also evaluate the overall ordering of recommendations using the area under the ROC curve (AUC), normalised Discounted Cumulative Gain (nDCG), and Mean Reciprocal Rank (MRR). The formula for computing AUC for a query $q$ given a list of relevant items $R$ is offered in Equation \ref{eq:auc} and can be interpreted as the likelihood that a random relevant item is recommended above a random irrelevant item.
\begin{equation}
\label{eq:auc}
    AUC(R)_q = \dfrac{1}{\vert R \vert (n-\vert R \vert)} \sum_{r \in R} \sum_{r' \notin R} \delta(r < r')
\end{equation}
Here $\delta(r < r')$ has the value $1$ if the relevant item $r$ is ranked above the irrelevant item $r'$ and $0$ otherwise. Similar to AUC, nDCG evaluates the ordering of the recommended items, but it includes an additional weighting to penalise cases where irrelevant items are ranked very highly.  Equation \ref{eq:dcg} describes the calculation of Discounted Cumulative gain with a binary definition of relevance.
\begin{equation}
\label{eq:dcg}
    DCG(R)_q = \sum_{i = 1}^{i\leq k} \dfrac{\delta(i)} {log_2(i+1)} 
\end{equation}
In this case $\delta(i)$ takes the value 1 if $i$ is a relevant item (i.e., $i \in R$) and 0 otherwise. To normalise DCG to yield a score in the range of [0,1], we divide by the ideal DCG score for the set of recommendations. MRR simply reports the reciprocal of the average position/rank of the first relevant item. For example, an MRR of 1.0 would indicate that the top recommendation was relevant, while a score 0.1 would indicate that the 10$^{th}$ recommended item was relevant. 
AUC, nDCG, and MRR scores are presented in Table \ref{tab:auc}.

In addition to recommendation relevance, we consider the recommendation novelty and recommendation diversity of the result sets. Tables \ref{tab:novelty_text}, \ref{tab:novelty_graph}, \ref{tab:diversity_text}, and \ref{tab:diversity_graph} report the novelty and diversity of recommendations papers, according to both the network and content perspectives as discussed in Section \ref{sec:method}. Recommendation novelty and recommendation diversity may have broad appeal across many RP-Rec-Sys applications. For instance, they are valuable in minimising redundant recommendations, as discussed in Section \ref{sec:diversity}. Yet, they are uniquely important to the use-case central to this work: promoting interdisciplinary research, reading and citation patterns using RP-Rec-Sys. The results of these experiments are presented in Section \ref{sec:results} below. Code for the experiments is available at \url{https://github.com/eoghancunn/facilitating_idr}.

\section{Results and Discussion}
\label{sec:results}
Table \ref{tab:prec_rec} presents the precision and recall results for the content-based recommendation evaluation described in Section \ref{sec:method}.  The GNN-based methods (\textit{GraphSAGE and ComBSAGE}) outperform \textit{SPECTER} in recommendation precision. While some of the differences are significant ($p<0.05$), they are modest. \textit{GraphSAGE} has the highest recommendation recall, with no significant different between \textit{SPECTER} and \textit{ComBSAGE} (at $\%95$ confidence). According to these results, we conclude that the state-of-the-art methods perform with similar precision, and given the improved recall achieved by the \textit{GraphSAGE} method, it should be preferred for use-cases like literature review. Since the TF-IDF baseline is not competitive with the other state-of-the-art methods, we will not consider the TF-IDF results further. It would be misleading to compare the recommendation diversity and novelty of two different sets of recommendations that have very different relevance scores, as the utility of diverse or novel recommendations is contingent on their relevance.


\begin{table}[h!]
\centering
\resizebox{\linewidth}{!}{
\setlength{\tabcolsep}{1pt}
\begin{tabular}{lrrrrrr}
\toprule
{} & \multicolumn{3}{c}{Precision ($\mu$ ± 2$\sigma$)} & \multicolumn{3}{c}{Recall ($\mu$ ± 2$\sigma$)} \\
{} & \multicolumn{1}{c}{@10} & \multicolumn{1}{c}{@20} & \multicolumn{1}{c}{@30} & \multicolumn{1}{c}{@10} & \multicolumn{1}{c}{@20} & \multicolumn{1}{c}{@30} \\
\midrule
TF-IDF &  0.037±0.002 &  0.039±0.002 &  0.040±0.002 &  0.002±0.000 &  0.003±0.000 &  0.005±0.001\\
GraphSAGE &  0.134±0.005 &  0.129±0.005 &  0.125±0.004 &  \textbf{0.018}±0.002 &  \textbf{0.029}±0.002 &  \textbf{0.041}±0.003 \\
SPECTER   &  0.126±0.005 &  0.117±0.004 &  0.113±0.004 &  0.014±0.002 &  0.023±0.002 &  0.031±0.002 \\
CombSAGE  &  \textbf{0.138}±0.005 &  \textbf{0.133}±0.005 &  \textbf{0.130}±0.005 &  0.012±0.001 &  0.022±0.002 &  0.031±0.002 \\
\bottomrule
\end{tabular}}
\vskip 0.5em
\caption{Precision and recall scores for 4 different paper embedding methods. Each method is used to produce the content representations for research papers in a content-based recommendation experiment. Every paper published in 2017 is given as a query to the system, and the model recommends the top 10, 20 and 30 papers. A recommendation is considered relevant if the query paper and recommended paper are co-cited at some point in the future (up to 2023). We compute precision and recall scores for all papers and use bootstrapping to report mean and standard deviation.}
\label{tab:prec_rec}
\end{table}

Table \ref{tab:auc} reports the AUC scores for recommendations offered by each embedding method. These scores show that throughout the entire ordering of recommended papers, the \textit{ComBSAGE} method is competitive with the state-of-the-art embedding methods when recommending relevant papers ahead of irrelevant papers. The nDCG scores confirm this result, although this metric shows that the \textit{GraphSAGE} method may be less likely to include irrelevant recommendations in the highest recommendations. Finally, the MRR scores measure the average position of the first relevant recommendation. According to these values, \textit{GraphSAGE, ComBSAGE} and \textit{SPECTER} offer their first relevant recommendation at positions 3.3, 3.8 and 4.3 respectively.

\begin{table}[h!]
\centering
\begin{tabular}{lrrr}
\toprule
 & AUC & nDCG & MRR \\
\midrule
GraphSage & 0.838 & \textbf{0.450} & \textbf{0.300} \\
SPECTER & 0.822 & 0.427 &  0.233 \\
ComBSAGE & \textbf{0.840} & 0.440 & 0.264 \\
\bottomrule
\end{tabular}
\vskip 0.5em
\caption{Area under the ROC curve (AUC), normalised Discounted Cumulative Gain (nDCG) and Mean Reciprocal Rank (MRR) for 3 different paper embedding methods. Each method is used to produce the content representations for research papers in a content-based recommendation experiment. Every paper published in 2017 is given as a query to the system, and all possible recommendations are ranked. A recommendation is considered relevant if the query paper and recommended paper are co-cited at some point in the future (up to 2023). We compute scores for all queries and report the mean.}
\label{tab:auc}
\end{table}

Tables \ref{tab:novelty_text} and \ref{tab:novelty_graph} report recommendation novelty scores. As discussed in Section \ref{sec:measuring_diversity_and_novelty}, recommendation novelty is measured as the average distance/dissimilarity between a query paper and a recommended paper, and is measured from two perspectives. Table \ref{tab:novelty_text} reports the \textit{article content} recommendation novelty scores measured using distances between \textit{SciBERT} embeddings, and Table \ref{tab:novelty_graph} reports \textit{citation network} recommendation novelty scores measured using distances between \textit{DeepWalk} embeddings. According to Table \ref{tab:novelty_text}, the textual content of recommendations made by the \textit{ComBSAGE} model are more distant from the query than those made by the \textit{GraphSAGE} or \textit{SPECTER} models. A similar pattern is also apparent in Table \ref{tab:novelty_graph}. That is, \textit{ComBSAGE} recommends research papers that are further from the query in the citation graph; these are more novel recommendations in the sense of Equation \ref{eq:novelty}. We also compute the novelty of the subset of the recommended papers that are \textit{relevant} to the query. Again we find that the \textit{ComBSAGE} model provides the most distant recommendations even when we consider only the relevant papers. 

\begin{table}[h!]
\centering
\setlength{\tabcolsep}{3pt}
\begin{tabular}{lrrrrrr}
\toprule
{} & \multicolumn{6}{c}{Novelty (Title+Abstract Distance)} \\
{} & \multicolumn{3}{c}{Recommended} & \multicolumn{3}{c}{(Relevant Subset)} \\
{} & \multicolumn{1}{c}{@10} & \multicolumn{1}{c}{@20} & \multicolumn{1}{c}{@30} & \multicolumn{1}{c}{@10} & \multicolumn{1}{c}{@20} & \multicolumn{1}{c}{@30} \\
\midrule
GraphSAGE &                  0.28&  0.28&  0.28&             0.20&  0.20&  0.20\\
SPECTER   &                  0.28 &  0.28 &  0.28 &             0.22 &  0.22 &  0.22 \\
CombSAGE  &                  \textbf{0.31} &  \textbf{0.30} &  \textbf{0.30} &             \textbf{0.25} &  \textbf{0.25} &  \textbf{0.25} \\
\bottomrule
\end{tabular}
\vskip 0.5em
\caption{Novelty of recommendations according to \textit{SciBERT} embeddings of query and recommended papers. At $k = 10$, the increase in novelty of the \textit{relevant} recommendations offered by \textit{CombSAGE} (according to the Cohen's $d$ effect size), is `\textit{moderate}' when compared to the next most precise method, \textit{GraphSAGE} ($d = 0.42$), and `\textit{small}' when compared to \textit{SPECTER} ($d = 0.2$).}
\label{tab:novelty_text}
\end{table}

\begin{table}[h!]
\centering
\setlength{\tabcolsep}{3pt}
\begin{tabular}{lrrrrrr}
\toprule
{} & \multicolumn{6}{c}{Novelty (Graph Distance)} \\
{} & \multicolumn{3}{c}{Recommended} & \multicolumn{3}{c}{(Relevant Subset)} \\
{} & \multicolumn{1}{c}{@10} & \multicolumn{1}{c}{@20} & \multicolumn{1}{c}{@30} & \multicolumn{1}{c}{@10} & \multicolumn{1}{c}{@20} & \multicolumn{1}{c}{@30} \\
\midrule
GraphSAGE &               0.55 &  0.55 &  0.55 &             0.37 &  0.37 &  0.37 \\
SPECTER   &               0.58 &  \textbf{0.58} &  \textbf{0.58} &             0.42 &  0.42 &  0.42 \\
CombSAGE  &               \textbf{0.59} &  \textbf{0.58} &  \textbf{0.58} &             \textbf{0.46} &  \textbf{0.45} &  \textbf{0.45} \\
\bottomrule
\end{tabular}
\vskip 0.5em
\caption{Novelty of recommendations according to \textit{DeepWalk} embeddings of query and recommended papers. At $k = 10$, the increase in novelty of the \textit{relevant} recommendations offered by \textit{CombSAGE} (according to the Cohen's $d$ effect size), is `\textit{moderate}' when compared to the next most precise method, \textit{GraphSAGE} ($d = 0.43$), and `\textit{small}' when compared to \textit{SPECTER} ($d = 0.07$).}
\label{tab:novelty_graph}
\end{table}

To help contextualise the benefits of the proposed \textit{ComBSAGE} embedding method, consider the following: at $k = 10$, the average \textit{relevant} recommendation made by the model employing \textit{ComBSAGE} has a \textit{DeepWalk} distance score of 0.46 when compared with the query paper. On average, in the citation graph, this corresponds to a shortest path distance of 2.2 citations. The next most precise model employs the \textit{GraphSAGE} method. In this case, the corresponding \textit{DeepWalk} distance score is 0.37, representing a shortest path distance of 1.7 citations on average. According to Cohen's $d$ effect sizes, the increases in novelty (or `unexpectedness') achieved by the \textit{CombSAGE} model range from \textit{small} to \textit{moderate} (see table captions). And, crucially, these increases come at no cost to precision. 

Finally, Tables \ref{tab:diversity_text} and \ref{tab:diversity_graph} report the recommendation diversity, measured as the average pairwise distance/dissimilarity between recommended papers, given the \textit{SciBERT} embeddings and \textit{DeepWalk} embeddings for papers respectively.
According to Table \ref{tab:diversity_text} (\textit{article content} recommendation diversity), the set of recommendations made by the \textit{ComBSAGE} model are more semantically diverse than those made by either the \textit{GraphSAGE} or \textit{SPECTER} models. Similarly, \textit{ComBSAGE} recommends research papers that are distributed more widely across the full citation graph (see Table \ref{tab:diversity_graph}, \textit{citation network} recommendation diversity). Although we calculate and include the diversity scores for the subset of recommendations that are \textit{relevant} for completeness, those scores are calculated on small sets of papers and thus may not be informative. Again, we find that effects according to Cohen's $d$ (see table captions) range from \textit{small} to \textit{moderate}.

\begin{table}[h!]
\centering
\setlength{\tabcolsep}{3pt}
\begin{tabular}{lrrrrrr}
\toprule
{} & \multicolumn{6}{c}{Diversity (Title+Abstract Distance)} \\
{} & \multicolumn{3}{c}{Recommended} & \multicolumn{3}{c}{(Relevant Subset)} \\
{} & \multicolumn{1}{c}{@10} & \multicolumn{1}{c}{@20} & \multicolumn{1}{c}{@30} & \multicolumn{1}{c}{@10} & \multicolumn{1}{c}{@20} & \multicolumn{1}{c}{@30} \\
\midrule
GraphSAGE &                    0.26 &  0.27 &  0.27 &             0.10 &  0.12 &  0.14 \\
SPECTER   &                    0.25 &  0.26 &  0.27 &             0.10 &  0.12 &  0.13 \\
CombSAGE  &                    \textbf{0.27} &  \textbf{0.28} &  \textbf{0.29} &             \textbf{0.11} &  \textbf{0.13} &  \textbf{0.15} \\
\bottomrule
\end{tabular}
\vskip 0.5em
\caption{Diversity of recommendations according to \textit{SciBERT} embeddings of recommended papers. At $k = 10$, the increase in diversity of the recommendations offered by \textit{CombSAGE} (according to the Cohen's $d$ effect size), is `\textit{small}' when compared to the next most precise method, \textit{GraphSAGE} ($d = 0.27$), and `\textit{moderate}' when compared to \textit{SPECTER} ($d = 0.5$).}
\label{tab:diversity_text}
\end{table}

\begin{table}[h!]
\centering
\setlength{\tabcolsep}{3pt}
\begin{tabular}{lrrrrrr}
\toprule
{} & \multicolumn{6}{c}{Diversity (Graph Distance)} \\
{} & \multicolumn{3}{c}{Recommended} & \multicolumn{3}{c}{(Relevant Subset)} \\
{} & \multicolumn{1}{c}{@10} & \multicolumn{1}{c}{@20} & \multicolumn{1}{c}{@30} & \multicolumn{1}{c}{@10} & \multicolumn{1}{c}{@20} & \multicolumn{1}{c}{@30} \\
\midrule
GraphSAGE &                 0.51 &  0.53 &  0.55 &             0.18 &  0.22 &  0.25 \\
SPECTER   &                 0.52 &  0.56 &  0.57 &             0.19 &  0.23 &  0.26 \\
CombSAGE  &                 \textbf{0.55} &  \textbf{0.58} &  \textbf{0.59} &             \textbf{0.21} &  \textbf{0.25} &  \textbf{0.28} \\
\bottomrule
\end{tabular}
\vskip 0.5em
\caption{Diversity of recommendations according to \textit{DeepWalk} embeddings of recommended papers. At $k = 10$, the increase in diversity of the recommendations offered by \textit{CombSAGE} (according to the Cohen's $d$ effect size), is `\textit{small--moderate}' when compared to the next most precise method, \textit{GraphSAGE} ($d = 0.31$), and `\textit{small}' when compared to \textit{SPECTER} ($d = 0.09$).}
\label{tab:diversity_graph}
\end{table}


\section{Conclusions}
\label{sec:conclusions}

The audiences for RP-Rec-Sys and their corresponding use-cases are numerous and varied. Each distinct use-case may require a different approach to recommendation to address different end-user needs. Despite this, a recent survey underscores that many studies have overlooked specifying a target audience or use-case for their proposed system \citep{kreutz2022scientific}. In this work, we propose a novel approach to RP-Rec-Sys evaluation. This approach aims to provide a more comprehensive characterisation of the recommendation outputs, allowing for the assessment of a given system with respect to its intended use-case. Specifically, our evaluation focuses on the \textit{diversity} and \textit{novelty} of research paper recommendations. Despite the attention and value ascribed to these metrics in the broader recommender systems literature \citep{smyth2001similarity,castells2021novelty}, they have been largely overlooked in RP-Rec-Sys contexts \citep{ali2021overview,kreutz2022scientific}.

Diverse and novel recommendations are desirable for many reasons, (see Section \ref{sec:background}), but for RP-Rec-Sys systems, they are uniquely important. Research paper recommendations that are novel and/or diverse can bridge different disciplines, helping to mitigate filter-bubble effects and siloed reading habits. They can actively promote interdisciplinary research, by facilitating the discovery of latent connections between otherwise distant topics, with the benefits of such interdisciplinary research being well-documented in \citep{shi2023surprising}. In pursuit of this goal, we have evaluated 4 different research paper representation methods in an RP-Rec-Sys. 

We have demonstrated how a recent approach (\textit{ComBSAGE}), can produce more more diverse and novel recommendations without compromising precision, when compared to other state-of-the-art methods. Thus we show the potential for RP-Rec-Sys to facilitate interdisciplinary knowledge transfer and prevent the perpetuation of information silos. Additionally, we can characterise the nature of recommendations provided by alternative methods, such that they might be prescribed to the use-case for which they are best suited. For example, with the highest recall, particularly at larger recommendation sizes ($k = 30$), the \textit{GraphSAGE} method may be the most appropriate choice of embedding method for scholars conducting literature review, In this scenario, a high recall on the relevant documents is likely to be paramount, whereas diverse and unexpected recommendations are potentially less useful. In this way, we recognise the measurable impact of document embedding methods on the quality and nature of downstream recommendations. Thus, we have demonstrated that the choice of representational approach within RP-Rec-Sys is important and should be grounded in the system's intended use-case or target audience.


\bibliography{references}

\section{Acknowledgments}
This research was supported by Science Foundation Ireland (SFI) under Grant Number SFI/12/RC/2289\_P2. 
This work builds substantially on a short paper previously published in workshop proceedings \citep{cunningham2023graph}. 


\end{document}